\documentclass[usenatbib, useAMS]{mn2e}
\usepackage{times}
\usepackage{color}
\usepackage{multirow}
\usepackage{graphicx}
\usepackage{amsmath}
\usepackage{natbib}
\usepackage{myaasmacros}
\usepackage{enumerate}
\usepackage{bm}

\usepackage[english]{babel}
\usepackage{mathptmx}
\usepackage{tabularx}
\usepackage{feynmp}
\usepackage{caption}
\usepackage{float}

\newcommand{\Tbub}{T_{\mathrm{bub}}}
\newcommand{\Tbase}{T_{\mathrm{base}}}
\newcommand{\rhobase}{\rho_{\mathrm{base}}}

\newcommand{\Msol}{\mathrm{M}_{\odot}}
\newcommand{\kpc}{\mathrm{kpc}}

\newcommand{\s}{\mathrm{s}}
\newcommand{\yr}{\mathrm{yr}}
\newcommand{\km}{\mathrm{km}}

\def\lsim{\mathrel{\rlap{\lower3pt\hbox{$\sim$}}
    \raise1pt\hbox{$<$}}}                
\def\gsim{\mathrel{\rlap{\lower3pt\hbox{$\sim$}}
    \raise1pt\hbox{$>$}}}                

\newcommand{\cm}{\mathrm{cm}}
\newcommand{\Myr}{\mathrm{Myr}}

\newcommand{\Gyr}{\mathrm{Gyr}}
\newcommand{\ergs}{\mathrm{ergs}}
\newcommand{\tcyc}{t_{\mathrm{cyc}}}
\newcommand{\fcyc}{f_{\mathrm{cyc}}}
\renewcommand{\vec}[1]{\bm{#1}}
\newcommand{\K}{\mathrm{K}}

\begin{document}
\pubyear{2015}
\title [Bursty star formation and cooling outflows]{Bursty star formation feedback and cooling outflows} \author[Suarez et al]{Teresita Suarez$^1$, Andrew Pontzen$^1$, Hiranya V. Peiris$^1$, Adrianne Slyz$^2$ \and and Julien Devriendt$^2$
  \\
  $^1${Department of Physics and Astronomy, University College
    London,
    London WC1E 6BT} \\
  $^2${Oxford Astrophysics, Denys Wilkinson Building, Keble Road, Oxford, OX1 3RH} \\
 }

\date{ Received ---; published---. }
\maketitle

\begin{abstract}
We study how outflows of gas launched from a central galaxy undergoing repeated starbursts propagate through the circumgalactic medium (CGM), using the simulation code {\sc Ramses}.  We assume that the outflow from the disk can be modelled as a rapidly moving bubble of hot gas at $\mathrm{\sim1\;kpc}$ above disk, then ask what happens as it moves out further into the halo around the galaxy on $\mathrm{\sim 100\;kpc}$ scales. { To do this we run 60 two-dimensional simulations scanning over parameters of the outflow. Each of these is repeated with and without radiative cooling, assuming a primordial gas composition to give a lower bound on the importance of cooling.} In a large fraction of { radiative-cooling} cases we are able to form rapidly outflowing cool gas from in situ cooling of the flow. We show that the amount of cool gas formed depends strongly on the `burstiness' of energy injection; sharper, stronger bursts typically lead to a larger fraction of cool gas forming in the outflow. The abundance ratio of ions in the CGM may therefore change in response to the detailed historical pattern of star formation. For instance, outflows generated by star formation with short, intense bursts contain up to 60 per cent of their gas mass at temperatures $<5 \times 10^4\,\mathrm{K}$; for near-continuous star formation the figure is $\lsim$ 5 per cent. Further study of cosmological simulations, { and of idealised simulations with e.g., metal-cooling, magnetic fields and/or thermal conduction}, will help to understand the precise signature of bursty outflows on observed ion abundances.
\end{abstract}

\vspace{1.5cm}

\section{Introduction}

Star-forming galaxies are surrounded by gas (known as the circumgalactic medium, CGM) with a comparable total mass to their stellar mass \citep{2014ApJ...792....8W}. This gas is enriched by metals that were almost certainly ejected from the galaxy; outflows, carrying a mass comparable to the mass of star-forming regions, are ubiquitously observed  in star forming galaxies in the local universe \citep{2004ApJ...606..829S,2015Natur.523..169E}, at intermediate redshifts $\lowercase { {z \sim 0.5}}$ \citep{Rubin2010, Nielsen2013, Bordoloi2014} and at high redshifts of $\lowercase{z \sim 6}$ \citep{2002ApJ...576L..25A,2005ApJ...621..227M}. 
The interrelationship between inflow and outflow is critical to the behaviour of galaxies as a whole, as it reshapes quantities such as the stellar mass function \citep{1986ApJ...303...39D,2010MNRAS.406.2325O} and mass-metallicity relation \citep{2008MNRAS.385.2181F}. Baryon cycling through the CGM likely also plays a role in controlling the distribution of dark matter \citep{pontzen,2015MNRAS.451.1366P}. 

The mechanism behind outflows is uncertain and may relate to some combination of supernova feedback \citep{2010Natur.463..203G}, winds of high-mass stars \citep{1999ApJ...513..156M, 2008MNRAS.387.1431D} and active galactic nuclei (AGN)  \citep{2008A&A...491..407N}. 
Dwarf galaxies are of particular interest since their small black hole masses makes AGN feedback ineffective, so that the outflows are almost certainly linked directly to star formation feedback. Additionally the higher gas-to-stellar-mass fraction combined with the shallow gravitational potential allow outflows to easily release material to the CGM \citep{2014ApJ...786...54P}. Observations in this low-mass regime
have shown evidence of C {\sevensize IV} absorption in the CGM out to 100 kpc.
Therefore, these galaxies are useful case studies of the connection between the CGM and the host galaxy.

Galactic outflows seem to possess a multi-phase nature, spanning several orders of magnitude in temperature \citep{2014ApJ...792....8W}. There is a particular puzzle over how cold gas material could survive in galactic outflows if they are entrained in a hot flow \citep[e.g.,][]{2005ARA&A..43..769V,2015ApJ...805..158S,2015arXiv150701951Z}. Perhaps this implies that cold gas can be directly accelerated using radiation pressure \citep{2005ApJ...618..569M,2012MNRAS.421.3522H}.  { The lifetime of cold clouds is, however, dependent on physical assumptions and simulation methods meaning that robust conclusions are difficult to draw. } Another scenario is that the cooler phases of the outflow are actually formed in situ by radiative cooling \cite[e.g.,][]{2000MNRAS.317..697E,2015ApJ...803....6M}; recently, detailed analytic discussions in support of this possibility have been given by \cite{2016MNRAS.455.1830T} and \cite{2015arXiv150907130B}.

In this paper, we study the formation of cool gas in outflows in a different way. Instead of developing analytic solutions to the outflow problem we inject hot gas into a idealised galaxy halo with a fixed potential. The numerical setup is similar to that of \cite{1999ApJ...513..142M} but introduces a simple prescription to deliver the outflow in discrete bursts. While our setup is highly simplified, we use it to argue that the balance between hot and cooler gas will be affected by the duty cycle of star formation. 
Such a link introduces a new dimension to the relationship between outflows and the evolution of a galaxy, raising the possibility that the relative abundance of different observed ions reflects information about the detailed star formation history. Since it is specifically {\it bursty} star formation that can have a profound impact on dark matter and stellar dynamics \citep{pontzen}, cross-checking typical star formation patterns in the observed universe would be a valuable additional benefit to studies of the CGM. { The simulations are two-dimensional, allowing us to run a much larger parameter study than would be possible in a three-dimensional study. }

This paper is organised as follows. In Sec.~\ref{simulations} we explain the initial and boundary conditions to simulate outflows in galaxies using {\sc Ramses} \citep{Teyssier:2002aa}. We discuss the results in Sec.~\ref{results}. Finally, we summarise in Sec.~\ref{conclusions}. 

\section{Hydrodynamical Simulations}
\label{simulations}

Our aim is to run a series of simulations that track how gas ejected from the disk of a galaxy interacts with halo gas out to the virial radius using {\sc Ramses}, a tree-based adaptive mesh refinement (AMR) hydrodynamical code.  We set up an idealised equilibrium halo, then inject hot outflows at the bottom of the computational box according to different models. To allow us to probe a large number of different scenarios we use 2-dimensional simulations.

\begin{figure}
    \centering
    \includegraphics[width=0.5\textwidth]{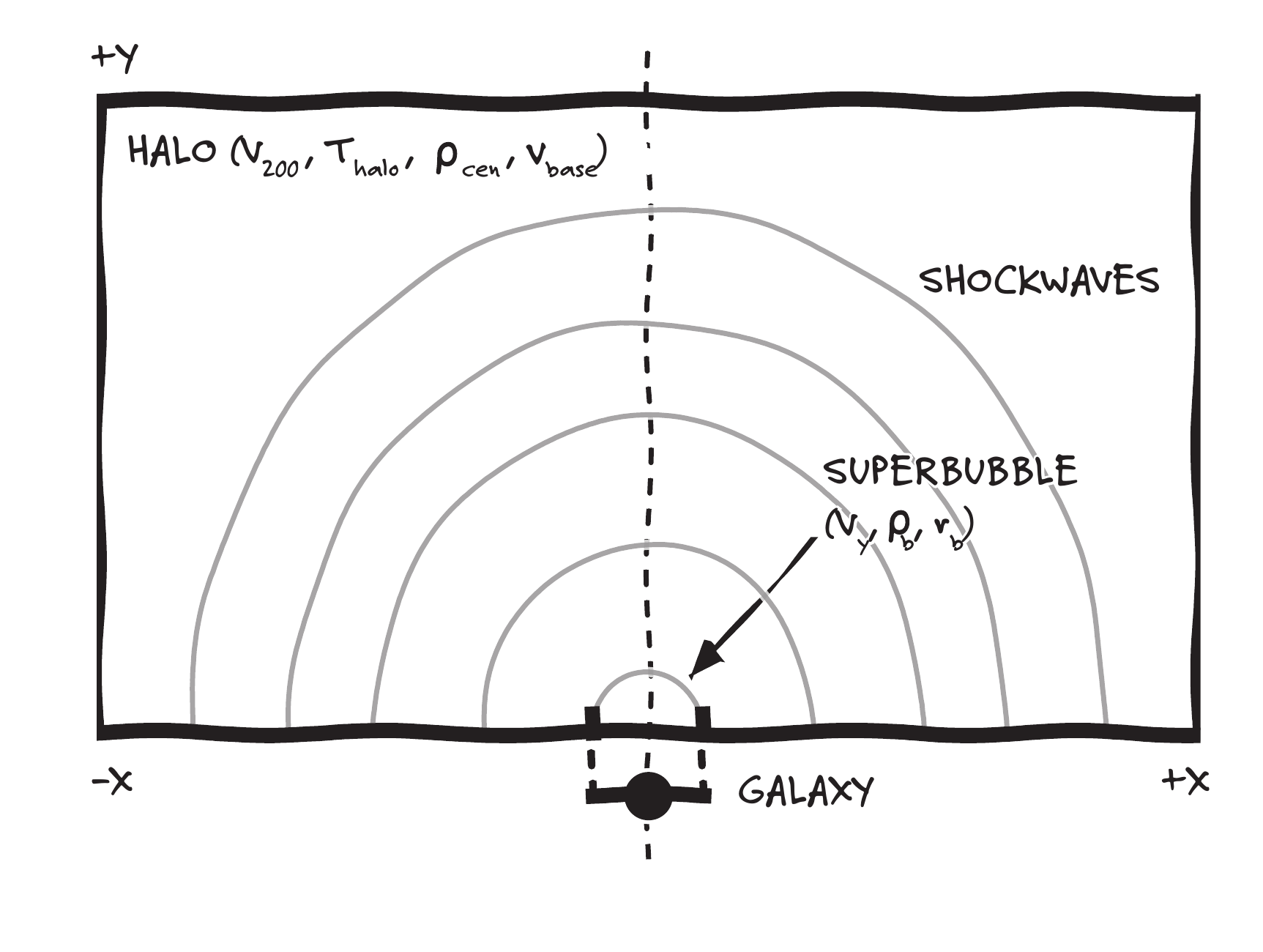}
    \centering
    \caption{A schematic representation of the two-dimensional set-up of our simulation. In our model, gas is assumed to flow into the box along the lower edge i.e., at $\mathrm{y=0}$, and flow out of the box along each of the remaining three edges.
We model half the halo with a radius of approximately $\mathrm{100\;kpc}$. The superbubble is referred to a starburst-type galaxy that is not part of the simulation. The physical properties of the gas expelled by the galaxy determine the boundary conditions in our simulations such as $\mathrm{r_{b}\sim 4\;kpc}$, $\mathrm{v_y \sim 400\;km\,s^{-1}}$, and a density $\mathrm{\rho_{b} \sim 0.05 \rho_{cen}}$, as well as the ghost regions that delimit the box. The physical properties of the halo such as $\mathrm{v_{200},\,T_{halo},\,\rho_{cen}\, and \, v_{base}}$ determine the initial conditions.}
    \label{boundary_conditions}
\end{figure}

In Fig.~\ref{boundary_conditions} we show a schematic representation of the CGM box in {\sc Ramses}. The galaxy is notionally positioned $d_{\mathrm{gal}}=1\,\mathrm{kpc}$ below the bottom of the 
box and it is not part of our simulations. Instead, we  assume hot gas from supernovae and other stellar feedback processes is expelled by the galaxy according to parameters that are described in Sec.~\ref{sec:boundary-conditions}.
The lateral and top sides of the box are chosen to satisfy {\it outflow} conditions, meaning that {\sc Ramses} sets gradients across them to zero. This way we allow gas to flow outside the box into the intergalactic medium (IGM).

For simplicity of setting up an equilibrium halo (which we discuss further in Sec.~\ref{sec:initial-conditions}), we assume a plane-parallel configuration so that the gravity is everywhere in a downwards direction. We assume a fixed gravitational field $g$ corresponding to that of a Navarro$-$Frenk$-$White (NFW) dark matter halo, but directed along the $\hat{\vec{y}}$ direction. This gives us a force law
\begin{equation}
g = - \frac{4 \pi G \rho_0 r_s^3}{y^2} \left[ \log \left(\frac{y+r_s}{r_s}\right) - \frac{y}{y+r_s} \right],\label{eq:nfw-g}
\end{equation}
where $\rho_0$ and $r_s$ are the scale density and radius of the NFW profile respectively. It will help to define the virial radius $r_{200}$ which is the radius containing a mean density 200 times that of the cosmological critical value. Our focus is on dwarf galaxies for the reasons outlined in the introduction; accordingly we use a halo with virial velocity $v_{200} = 50\,\mathrm{km\, s}^{-1}$, giving a virial radius $r_{200} \simeq 140(1+z)^{-3/2}\,\mathrm{kpc}$ and virial mass $M_{200} \simeq 8 \times 10^{10}\,(1+z)^{-3/2}\,\Msol$.  We adopt the scale radius $r_s = 5$ kpc from the fit given by \cite{2007MNRAS.378...55M}.

The mesh in {\sc Ramses} is defined on a recursively refined spatial tree. We set the maximum level of refinement to 10, which means that the regular Cartesian grid is continuously refined in the course of the simulation by a factor of up to $2^{10}=1024$; our box size of $80\,\mathrm{kpc}$ therefore allows for a maximum resolution of $80\,\mathrm{pc}$. We ran a convergence test with higher resolution on one case, reaching $40\,\mathrm{pc}$ by increasing the maximum refinement to 11, finding that the results that we describe below did not change. The refinement strategy opens a new cell if the discontinuity in density or pressure is above $5$ per cent; for time stepping we adopted a Courant number of 0.6.

\subsection{Initial conditions: equilibrium inflow}\label{sec:initial-conditions}
We modified the {\sc Ramses} code in order to set up equilibrium, inflowing (or hydrostatic) gas in our fixed potential to represent the halo and to inject hot gas in the bottom. Here we discuss the initial equilibrium, then discuss the injection method in Sec.~\ref{sec:boundary-conditions}.

We ran simulations both with and without cooling. While a hydrostatic solution is an attractive initial stable state in terms of its simplicity, there are no such solutions when cooling is enabled. Therefore we always consider inflowing gas. 
The first fluid motion equation enforces mass conservation so that for any region the rate of change of its mass is the net flow of mass into it:
\begin{equation}
{\frac{\partial \rho}{\partial t} + \rho \frac{\partial u}{\partial y} + u \frac{\partial \rho}{\partial y}=0 \textrm{,}}\\
\label{massconservation}
 \end{equation}  
where $\rho(y)$ is the density, $u(y)$ is the velocity in the $y$ direction, and $t$ denotes time. The momentum-conservation equation ensures the rate of change of momentum is balanced by momentum flow and net force:
\begin{equation}
\rho \frac{\partial { u}}{\partial t} + \rho u \frac{\partial u}{\partial y}= - \frac{\partial p}{\partial y} -\rho g\textrm{,}\\
\label{momentumeqn}
 \end{equation}  
\noindent where $p$ is the pressure. Finally we have the energy conservation equation:
\begin{equation}
{\frac{\partial E}{\partial t}+ \frac{\partial}{\partial y} [(E+p){u}]=-\rho \dot Q_{\mathrm{cool}}+\rho \frac{\partial \Phi}{\partial t}}\textrm{,}\\
\label{energyeqn}
\end{equation}  
\noindent where  $E = \rho(\epsilon + u^2/2 + \Phi)$ refers to the energy per unit volume, $\epsilon$ is the internal energy per unit mass, $\Phi$ is the potential implied by Eq.~\eqref{eq:nfw-g}, and the cooling function is defined by  $\dot Q_{\mathrm{cool}}(\rho, T)$ per unit mass. We substitute for the internal energy
\begin{equation}
\epsilon = \frac{3 kT}{2 \mu m_p} \textrm{,}
\end{equation}
where $k$ is the Boltzmann constant, $m_p$ is the proton mass, and $\mu$ is the mean molecular weight of the gas constituents. We assume a primordial gas for our background solution, and furthermore that $\mu$ is constant (i.e., the ionisation is fixed), and check that this assumption holds by ensuring the resulting initial conditions are stable over many sound-crossing times.

\begin{figure}
    \centering
    \includegraphics[width=0.45\textwidth]{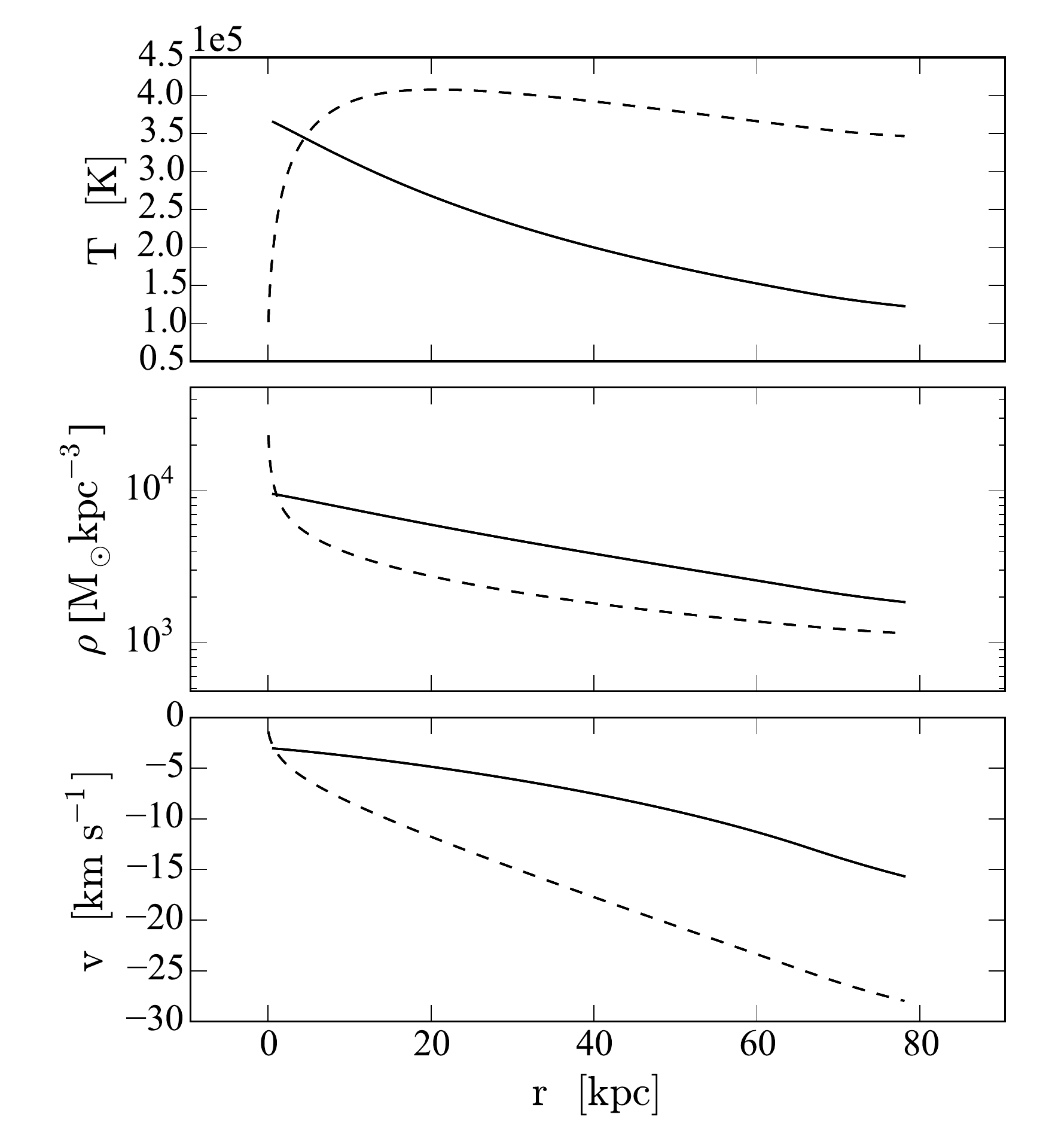}
    \centering
    \caption{Equilibrium inflow conditions that we adopted as a function of height. These are obtained by solving Eqs. \eqref{eqn:u}, \eqref{eqn:rho} and \eqref{eqn:Temp}. The solid line represents the adopted solution for adiabatic runs (i.e., ${\dot Q_{\mathrm{cool}}=0}$). The solution when cooling is activated is represented by the dashed lines. }
    \label{big_galaxy_properties}
\end{figure}

To obtain the equilibrium halo solution, we then set all partial derivatives with respect to $t$ to be zero. Overall, once simplified using Eq.~\eqref{massconservation}, the energy equation, Eq.~\eqref{energyeqn}, becomes
\begin{equation}
u \frac{\partial u}{\partial y} + \frac{5 k}{2 \mu m_p} \frac{\partial T}{\partial y} = g - \frac{\dot Q_{\mathrm{cool}}}{u}\textrm{.}
\end{equation}
Finally we solve the three equations above for $\partial u/\partial y,\;\partial \rho/\partial y$ and $\partial T/\partial y$:

\begin{align} 
\frac{\partial u}{\partial y} & =\frac{ -2\dot Q_{\mathrm{cool}} +3 g u   }{-5\tilde{R}T+3u^2} \, , \label{eqn:u} \\
\frac{ \partial \rho}{\partial y} & = - \frac{\rho}{u} \frac{\partial u}{\partial y} \hspace{0.5cm} \textrm{ and } \label{eqn:rho} \\
\frac{\partial T}{\partial y} & = 2\; \frac{ (\dot Q_{\mathrm{cool}}/u) \tilde{R}T + g\tilde{R}T- \dot Q_{\mathrm{cool}}u }{k(-5\tilde{R}T+3u^2)}, \label{eqn:Temp}
\end{align} 
where $\tilde{R} = k / \mu m_p$. Note that the denominator can cross zero, corresponding to a shock in the inflow solution, but by studying only low mass galaxies we do not suffer from this potential problem \citep{1977MNRAS.179..541R}.  We used Eqs. \eqref{eqn:u} -- \eqref{eqn:Temp} in conjunction with a numerical Runge-Kutta integrator to obtain initial conditions. 

We required two solutions: one for the adiabatic case when we set ${\dot Q_{\mathrm{cool}} =0}$ and the other for cooling simulations. In the latter case we obtain $\dot Q_{\mathrm{cool}}$ from the {\sc Ramses} cooling function assuming primordial gas composition and a UV background fixed to the \cite{2012ApJ...746..125H} normalisation at $z=2$. At lower redshifts, metal enrichment and a reduced UV background would lead to faster cooling. Therefore our results will give a lower bracket for the amounts of cool gas that can be generated in outflows.

We have freedom in imposing conditions $\rho(0)$, $u(0)$ and $T(0)$ (i.e., the density, velocity and temperature $1\,\mathrm{kpc}$ above the notional galaxy). Our primary criteria for choosing these was to obtain a circumgalactic medium with (i) a density of around $10^4\;M_{\odot}\,\mathrm{kpc}^{-3}$ at $y=0$ reducing to around $10^3\;M_{\odot}\,\mathrm{kpc}^{-3}$ at the virial radius; (ii) an overall temperature of around $3 \times 10^5\;\mathrm{K}$; (iii) a small inflow velocity $\lsim 5\;\mathrm{km}\,\mathrm{s}^{-1}$ at the base of the box. These choices were motivated by the inflowing component of the circumgalactic medium of the cosmological zoom simulation DG1 from \cite{2012MNRAS.421.3464P}.

Achieving these goals requires a different solution for the adiabatic and cooling cases. Figure \ref{big_galaxy_properties} shows the two solutions as respectively a dashed and solid line, illustrating (from top to bottom) the temperature, density and velocity. We tuned by hand to find solutions as similar as possible between the two cases. The biggest difference between cooling and adiabatic cases is that, when cooling is enabled, the temperature drops rapidly as inflowing gas nears the bottom of the box. This also implies that the density increases rapidly in the same region. 

Overall we found two solutions with similar overall parameters for the undisturbed CGM, and were able to verify that these were indeed stable by running them for several Gyr in {\sc Ramses}.

\begin{table}
\begin{center}
\begin{tabular}{  r | c  c  }
  Quantity & Values (cooling) & Values (adiabatic) \\
\hline \rule{0pt}{2.5ex}
  ${T_{\mathrm{base}}}$   & $6\times10^4$ $\mathrm{K}$ & $3\times10^5$ $\mathrm{K}$ \\
  ${\rho_{\mathrm{base}}}$ & $10^{-3}\, \mathrm{amu}\, \cm^{-3} $ & $3 \times 10^{-4}\, {\mathrm{amu}\, \cm^{-3}}$ \\
	& = 2.5$\times10^4$ M$_{\odot}$ kpc$^{-3}$	&= 7.4$\times10^3$ M$_{\odot}$ kpc$^{-3}$	\\
  ${v_{\mathrm{base}}}$ & 2 $\km\,\s^{-1}$ & 3 $\km\,\s^{-1}$ \\
\rule{0pt}{3ex}
  ${r_{\mathrm{b}}}$ &  \multicolumn{2}{c}{$4\,\kpc$} \\
  ${\rho_{\mathrm{b}}/\rho_{\mathrm{base}}}$ &  \multicolumn{2}{c}{$0.02/\fcyc$ (maintaining $\dot M_{\mathrm{av}}$)} \\
  ${v_{\mathrm{y}}}$ & \multicolumn{2}{c}{$400\,\km\,\s^{-1}$}\\
\rule{0pt}{3ex}
  ${t_{\mathrm{cyc}} }$ & \multicolumn{2}{c}{Vary between $100$ and $3000\,\Myr$} \\
  ${f_{\mathrm{cyc}}}$ & \multicolumn{2}{c}{Vary between 0.033 and 1} \rule{0pt}{2.5ex} \\
  \hline
\end{tabular}
\end{center}
\caption{Parameters describing initial and boundary conditions for the models, motivated by matching onto existing ab initio galaxy simulations as explained in the text. The steady-state background inflow solutions are determined by the first three parameters and vary between cooling and non-cooling simulations to better match the mean properties of the halos. Instantaneous properties of the ``bubble'' emerging into the bottom of our box are given by the next three parameters. The final two parameters, $\fcyc$ and $\tcyc$, describe the cycle of outflow activity. To keep the time-averaged outflow rate $\dot M_{\mathrm{av}}$ the same between simulations, $\rho_b$ is allowed to depend on $\fcyc$.}
\label{on_off}
\end{table}

\begin{figure*}
    \centering
    \includegraphics[width=0.95\textwidth]{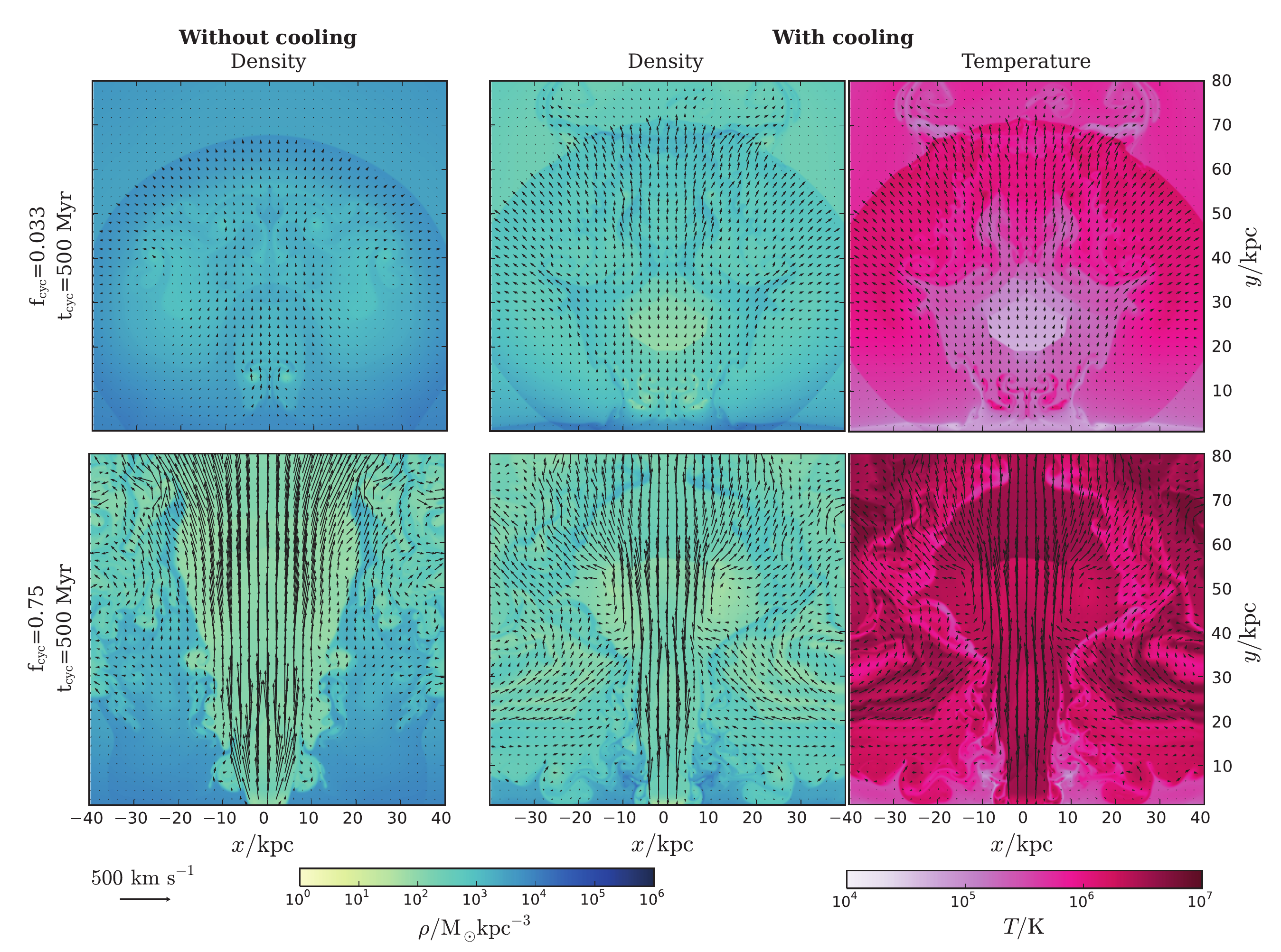}
    \centering
    \caption{Example renderings of our simulations after several Gyr. The top row shows a short duty cycle example with $\tcyc = 500\;\Myr$ and $\fcyc=0.033$. The bottom row shows a long duty cycle example with $\tcyc=500\;\Myr$ and $\fcyc = 0.75$. From left to right the panels show the density in the non-cooling (adiabatic) simulations, the density with cooling enabled, and the temperature with cooling enabled. }
    \label{illustration}
\end{figure*}
\subsection{Outflow characterisation}\label{sec:boundary-conditions}
In the previous section we described the physical properties of inflowing gas, setting our initial conditions. We now consider the boundary conditions which include the properties of the ``superbubble'' emerging from the galaxy, assuming a single phase outflow uniformly distributed across a radius of $r_{\mathrm{b}}$. 
The properties of the new hot material are characterised by the radius, the density of the new gas (${\rho_b}$) and its upwards velocity (${v_y}$).
The temperature of the superbubble is set for pressure equilibrium so that ${p_{\mathrm{b}}=p_{\mathrm{base}}}$, otherwise pressure would cause it to rapidly expand. Since ${p\propto\rho T}$, the temperature and density of the bubble and inflow are interrelated by ${\Tbub/\Tbase=\rhobase/\rho_b}$. We assume that the outflow is dominated by a hot phase with $\Tbub \gg \Tbase$, ignoring any colder material that may have been transported from the disk. While it may be possible to accelerate cold clouds, preventing their disruption over a few scale-lengths is challenging \citep{2015ApJ...805..158S} and consequently on large scales, in situ cooling is likely to be a strongly dominant source of cool outflows \citep{2016MNRAS.455.1830T}. 

The main aim of the current paper is to investigate how in situ cooling changes as the nature of the star formation cycle in the galaxy is altered. To allow us to parameterise this and investigate systematic changes, we define a {\it cycle length} $\tcyc$ and {\it duty cycle} $\fcyc$. The cycle length is the overall time periodicity of the galactic star formation.  The duration of the whole simulation is always set for at least five times the cycle length, which means that multiple bubbles are injected. The duty cycle is the fraction of time that the bubble actually spends pumping gas into the halo, so that $\fcyc=0$ implies no bubble is created whereas $\fcyc=1$ means that the galaxy expels gas at all times. Our cycle is deterministic; in this work we have not investigated the modifications that randomness within the duty cycle could introduce.

The two parameters $\tcyc$ and $\fcyc$ are of particular interest because they characterise a ``burstiness'' for the star formation; bursts of star formation are also important to determine whether dark matter cusp/core transformations are generated by the outflows \citep{pontzen}. In galaxy formation simulations, the burstiness of star formation is strongly affected by the feedback prescription \cite[e.g.,][]{2015arXiv150206371L,2015MNRAS.454.2092O}. We allow $\tcyc$ and $\fcyc$ to vary independently in respectively 8 and 5 steps, making a grid of 40 models. The values vary between $100$ and $3000\;\Myr$ for $\tcyc$ and between $0.033$ and $1.0$ for $\fcyc$.

When outflows are being generated, the amount of gas produced per unit time (${\dot M_{\mathrm{on}}}$) is determined by the velocity, the density and the radius of the bubble as 
\begin{equation}
{ \dot M_{\mathrm{on}} = 2 (\pi r_b^2)v_y \rho_b }\;\textrm{,}\\
\label{outflow_rate}
\end{equation}  
where the factor of two arises from assuming outflows to occur in both directions from the disk. In our scheme, the overall outflow rate is also dependent on the time that the galaxy is expelling gas, i.e., the averaged outflow rate is set by $\dot{M}_{\mathrm{av}} = {\dot M_{\mathrm{on}}\fcyc}$. The total amount of gas expelled by such galaxies is very uncertain, and, because it is too diffuse, halo gas is extremely difficult  to observe directly \citep{2014ApJ...792....8W}. We enforce in all our simulations that $\dot{M}_{\mathrm{av}} \simeq 0.01 \,\Msol\,\yr^{-1}$, motivated by mass-loading factors of order unity, coupled to low average star formation rates expected for systems of this size \citep{2010Natur.463..203G,2015arXiv150800007C}. Consequently $\dot M_{\mathrm{on}}$ must vary as $\fcyc$ changes in our simulations. Specifically there is the following relationship between our parameters:

\begin{equation}
\frac{\dot M_{\mathrm{av}}}{0.01\,\Msol\,\yr^{-1}} = \frac{v_y}{400\,\km\,\s^{-1}} \frac{\rho_b}{490\;\mathrm{M_{\odot}\,kpc^{-3}}} \left( \frac{r_b}{4\,\kpc} \right)^2 \fcyc \textrm{.}
\end{equation}

The default parameters here have again been motivated by a study of DG1. We fix the left-hand-side of this relationship but vary $\fcyc$, so are forced to change other parameters of the outflow. While there is no unique prescription for this, we chose to fix $v_y$ (so that the outflows are always strongly supersonic) and $r_b$ (because this is limited by the size of the disk), allowing $\rho_b$ to change $\propto \fcyc^{-1}$ in compensation. In our most extreme case of $\fcyc=0.033$, $\rho_b \simeq { 1.7 \times10^4 \; M_{\odot}\, kpc^{-3}} \simeq 7 \times 10^{-4}\,\mathrm{amu}\,\cm^{3} \simeq 0.7 \rhobase$. Note that there is no dependency on $\tcyc$, so that our models at fixed $\fcyc$ have identical outflow parameters.

Because the velocities are quite substantial, most of the energy of our bubbles is tied up in kinetic form (even though in the disk, where pressures are substantially higher, the energy may have been thermal). The total energy available from supernova (SN) explosions is dependent on the stellar initial mass function, but is approximately $10^{49}\,\ergs$ per solar mass of stars formed. The velocity of a wind driven by this energy source is 
\begin{equation}
v_y \simeq \sqrt{\epsilon/0.1 \eta} \times 300\,\km\,\s^{-1}\textrm{,}
\end{equation} 
where $\epsilon$ is the fraction of $10^{49}\,\ergs\,\Msol^{-1}$ that actually couples and $\eta$ is the mass-loading factor of the wind, i.e., $\eta=\dot{M}_{av}/\dot{M}_{\star}$ where $\dot{M}_{\star}$ is the star formation rate. Our chosen parameters are consistent in the sense that recent simulations find order-unity mass loading $\eta$, while adopting values of $\epsilon$ that are at least $0.1$ and often considerably more \citep{2010Natur.463..203G,2014MNRAS.445..581H}. As we will see in the next section, the typical outflow velocities that would be observed along a random sightline through the halo is normally considerably less than the imposed outflow velocity at the bottom of our box, since the energy is rapidly dispersed in the halo.

We summarise the values of both the initial and bubble conditions in Table \ref{on_off}.


\section{Results}
\label{results}

As we explained in Sec.~\ref{simulations} and Table \ref{on_off} above, our exploration is based on a $8 \times 5$ grid of parameters $\tcyc$ and $\fcyc$. Note that when $\fcyc=1$, the $\tcyc$ parameter has no effect, so we can eliminate seven of the simulations from the grid.
We allow each simulation to run for $>12\;\Gyr$ so that the behaviour is not sensitive to the precise details of the initial conditions, even in the cases with a long $\tcyc$.

We made an initial visual inspection to classify the morphology of the outflow, finding that the duty cycle causes the adiabatic runs to fall into distinct regimes. Some example outputs are illustrated in Fig.~\ref{illustration}. The top row shows the simulation with $\fcyc=0.033$, $\tcyc=0.5\,\Gyr$ whereas the lower row is generated from simulations with  $\fcyc=0.75$, $\tcyc=0.5\;\Gyr$. The simulations are pictured at $t=5.9\;\Gyr$, after a repeatable cycle has been established (see below).

In each row, the two left panels show log gas density $\rho$ over the range ${10 \,M_{\odot}\; \kpc^{-3}\le \rho \le 10^6\;M_{\odot}\, \kpc^{-3}}$ for the non-cooling and the cooling simulation. The final panel shows the temperature in the cooling case, scaled between ${10^4\; \mathrm{K}\le T \le 10^7\;\mathrm{K} }$.  Arrows indicate the velocity vectors of gas cells in the halo.

At any given time the morphology of all our simulations roughly falls into one of three categories:

 \begin{figure}
     \centering
     \includegraphics[width=0.5\textwidth]{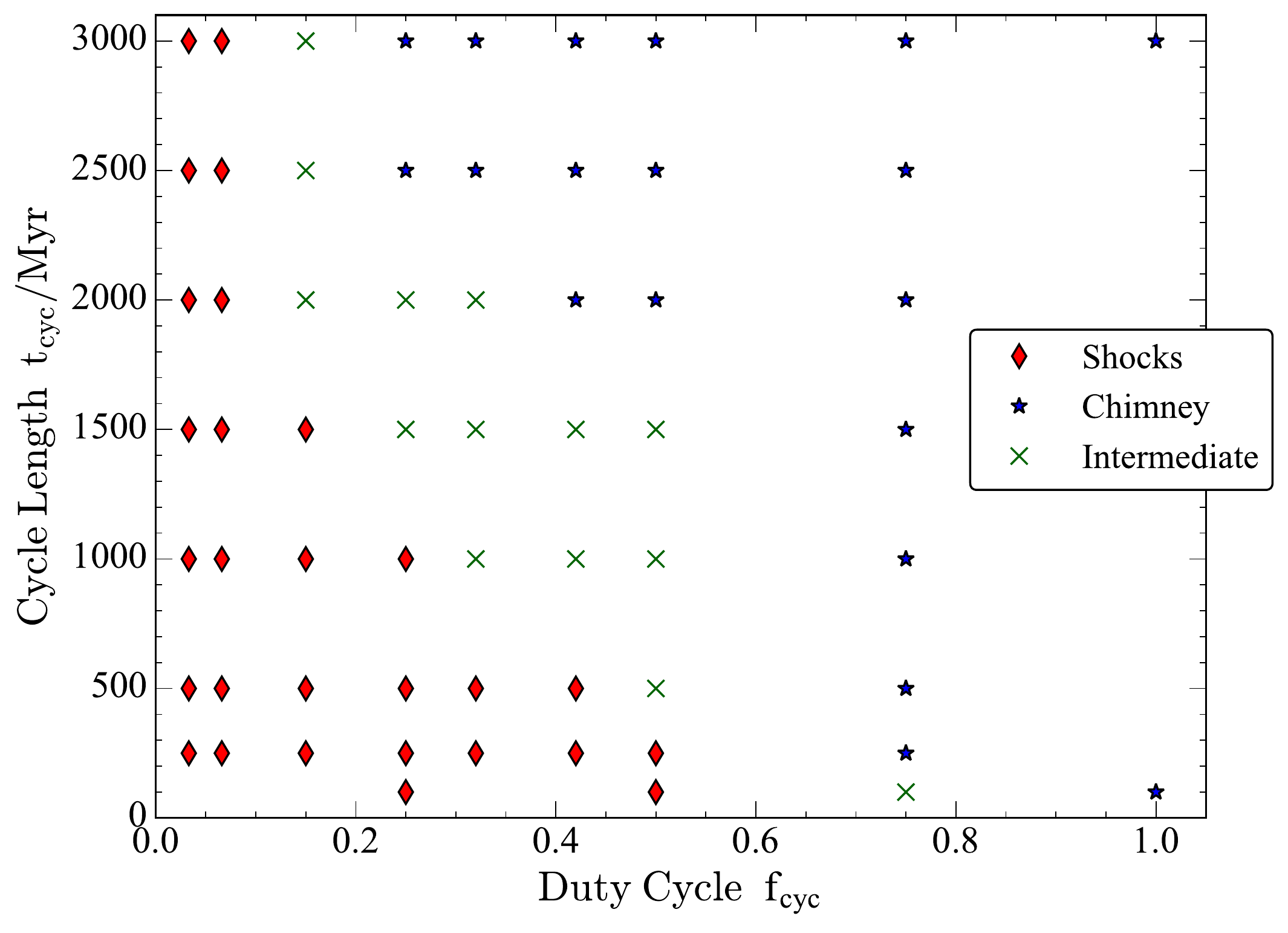}
     \centering
     \caption{Distribution of shock-dominated and chimney-dominated outflow type with the variations of duty cycle and cycle length parameters. We make this classification by visual inspection of the outflows as described in the text. Red diamonds, blue stars and green crosses indicate the shock-dominated, chimney-dominated and intermediate cases. We have used the simulations with cooling enabled, but found that the differences in the case without cooling were minor. 
{ To maintain the average mass-loading, the density of the bubble scales in inverse proportion to the duty cycle, ranging from  $1.7
\times 10^4\, \mathrm{M_{\odot}\,kpc^{-3}}$ for $\fcyc = 0.033$ to $6.1 \times 10^2\, \mathrm{M_{\odot}\,kpc^{-3}}$ for $\fcyc = 1.0$.}} 
\label{pattern}
\end{figure}

\begin{itemize}
\item{\it Shock-dominated},  where the morphology of the density and temperature maps is dominated by roughly spherical, discontinuous fluid flows (see e.g., top-left panel of Fig.~\ref{illustration}). These predominantly occur when $\tcyc$ is long and/or $\fcyc$ is small; the halo is able to settle down between each ``burst'' of wind injection which therefore triggers a significant shock wave. 

\item {\it Chimney-dominated,} with a morphology which is more dominated by a classic chimney- or funnel-shaped, relatively smooth flow of gas (see e.g., bottom-left panel of Fig.~\ref{illustration}). These can only be seen in our simulations with large $\fcyc$ or $\tcyc$, which allows the prolonged outflows to ``punch through'' the inflowing material. The halo material outside the main outflow often ends up being very turbulent in these cases.

\item {\it Intermediate} outflows have morphologies that share characteristics of the two types.
\end{itemize}

A single simulation can switch between the three phases above during its evolution. However typically shocks can only dominate for a limited time after the outflow switches on, after which it sweeps out of the halo into the IGM beyond our box. When $\fcyc$ is small (such as in the top row of Fig.~\ref{illustration}), the entire outflow consists of a series of outgoing shocks and a chimney never establishes itself. Conversely when $\fcyc$ is large, the outflow normally looks like a classic chimney pattern. Cases with intermediate $\fcyc$ values and large $\tcyc$ can switch between the two morphologies depending on the time of observation. 

As a rough illustration of this point, Fig.~\ref{pattern} shows our classification for the dominant morphology of each simulation. Shock-dominated outflows are represented with red diamonds and chimney-dominated cases with blue stars. Green crosses indicate the intermediate type where no clear classification can be decided on because it is strongly time-dependent.

\begin{figure}
    \centering
       \includegraphics[width=0.5\textwidth]{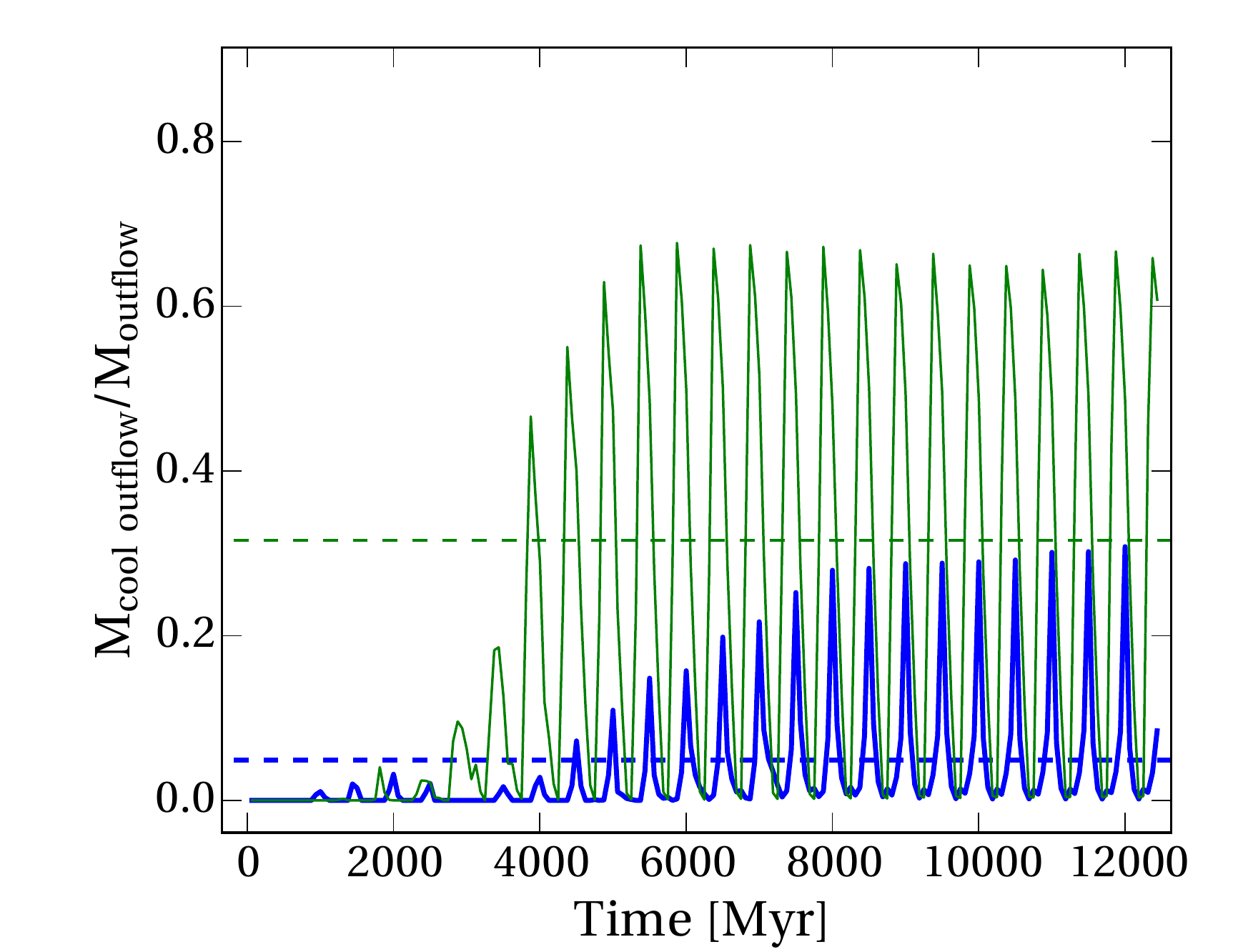}
    \centering
    \caption{Cool gas fraction comparison between the galaxies illustrated in Fig.~\ref{illustration}. For each timestep, the cool gas fraction is calculated as a mass fraction of outflowing gas ($v_y>20\,\km\,\s^{-1}$) that has temperatures satisfying $T<5 \times 10^{4}\,\K$. The thinner, green line shows the case $\fcyc=0.033$ whereas the thicker blue line shows the case $\fcyc=0.75$. The time-averaged value for the two cases is shown by corresponding dashed horizontal lines and is $32$ per cent and $5$ per cent for the two respective cases.}
    \label{coldgas}
\end{figure}

Comparing the upper-left panel and upper-mid panels of Fig.~\ref{illustration}, we can see that once cooling is enabled, the shocks have a different structure. The shell of dense material caused by the initial shock is able to cool efficiently and so the highest densities are formed behind the shock-front. In the illustrated snapshot, the shock from the most recent episode has reached around $70\;\kpc$ above the galaxy; above this are the remains of the previous shock sweeping through the halo. Behind the shock front, the dense gas reaches a regime where its radiative cooling time is short. The typical outflow speeds in the illustrated halo are $\sim 70\;\km\,\s^{-1}$; because the bursts of high-velocity outflows are so short, the initial energy is rapidly spread across a greater mass of gas, making for high mass loadings but lower velocities in the halo.

Conversely high-velocity cold gas is typically not generated in large quantities in simulations with a large $\fcyc$, where the chimney is the dominant morphology. The lower panels of Fig.~\ref{illustration} show how the gas establishes a direct route out of the halo and so maintains its high initial outflow speed once the initial resistance of the cooling inflow has been cleared away. This high speed, high temperature, relatively low-density flow rarely reaches a regime where it cools efficiently.

To make a more quantitative comparison between the cases and their ability to generate cool outflows, we quantified the fraction of the outflowing mass that is cool in each case. We defined outflowing gas as having an upwards vertical velocity $>20\,\km\,\s^{-1}$ and ``cool'' gas to be at temperatures $T<5\times 10^4\,\mathrm{K}$. For each snapshot we then measured the cool outflow fraction by mass. The results for the two simulations that we have discussed so far are shown in Fig.~\ref{coldgas}; the thicker and thinner lines respectively are generated from the chimney-dominated $\fcyc=0.75$ and shock-dominated $\fcyc=0.033$ cases.

Over the first few cycles, the amount of cold gas forming in both cases grows. Eventually a repeating cycle is established. Each cycle starts when the galaxy (below the box) starts injecting gas; at this point the amount of cool gas becomes small as the injected material is hot and compresses any cold material. Once the outflow shuts off, the fraction of cool gas tends to grow as the shock expands outwards and radiative effects become significant behind the shock front. Short $\fcyc$ values switch off the heating source earlier in the cycle and thus allow larger cool mass fractions to build up before the next cycle begins. This is reflected in the average, taken over the penultimate two cycles, of the cool mass fraction which is $32$ per cent and $5$ per cent in the $\fcyc=0.033$ and $\fcyc=0.75$ cases. These averages are shown by the horizontal dashed lines.

\begin{figure}
    \centering
    \includegraphics[width=0.5\textwidth]{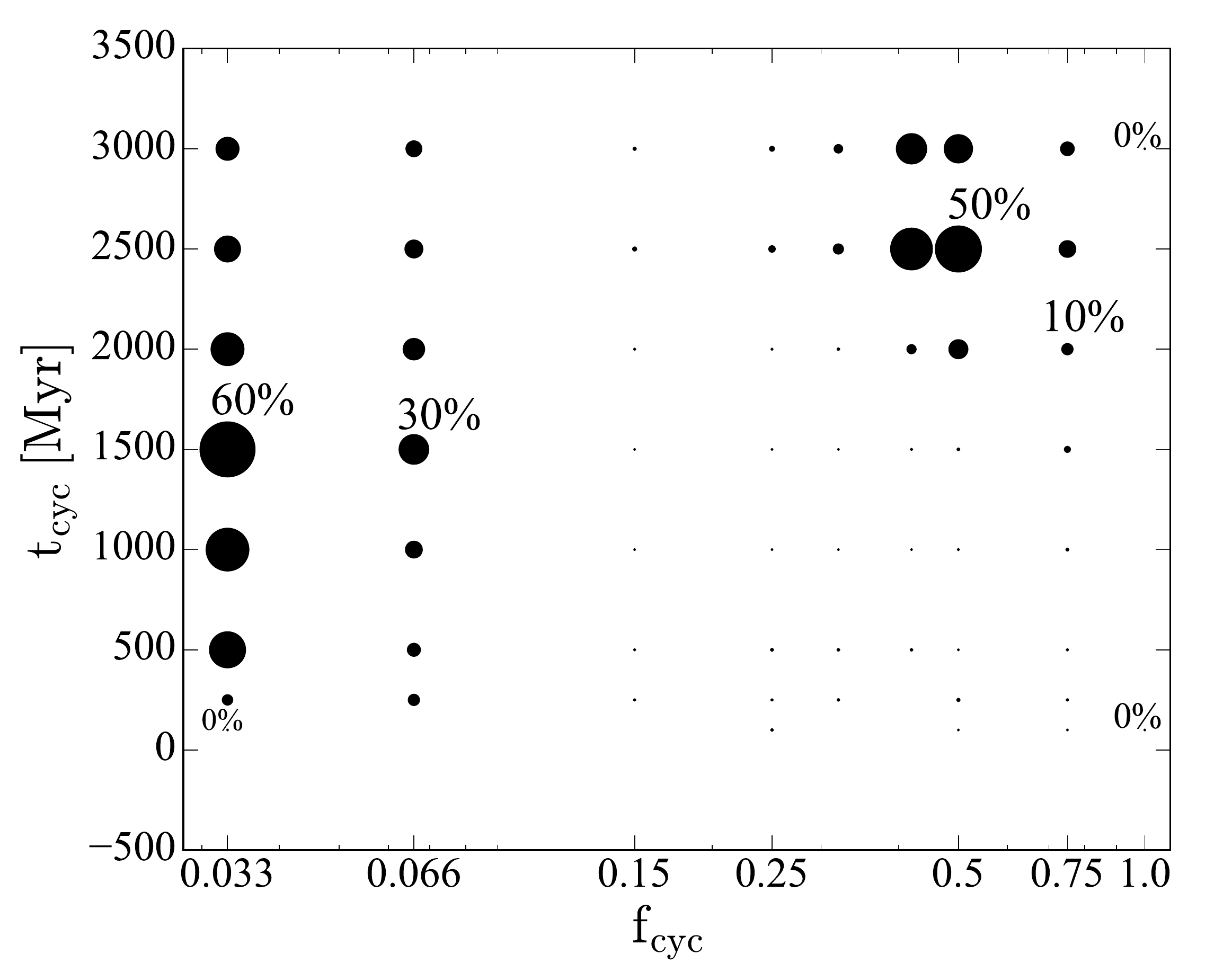}
    \centering
    \caption{Duty cycle ${f_{\mathrm{cyc}}}$  distribution with the cycle length, ${t_{\mathrm{cyc}}}$ as shown in Fig.~\ref{pattern} , but now the size of the circles show the amount of cool gas in the outflow { and the percentages show the numeric value for a selection of models.} 
{
Our results highlight two regimes in which cool gas is easiest to form; either a small duty fraction is required, giving relatively high outflow densities and so allowing gas to cool in between bursts, or the total cycle time must be long, yielding long time periods during which denser regions of the turbulent halo can cool.}
}
    \label{dutyvscycledots}
\end{figure}

In Fig.~\ref{dutyvscycledots} we show this average cool-mass-fraction of outflows calculated for each simulation. The position on the plot shows the duty cycle and cycle length, and the size of the plotted dot is in proportion to the cool-mass-fraction. The average is taken over gas outflowing at speeds greater than $\mathrm{50\;km\;s^{-2}}$ after $\mathrm{9\;Gyr}$, when the cycle has become stable. Our summary statistic of the total cool mass fraction gives a sense of how observable the cool phase is likely to be, but we emphasise that the cool gas often occupies a relatively small volume and is only present in large quantities during certain parts of the cycle. 

Outflows with our smallest duty cycle (i.e., on the left edge of the figure) are able to generate cool mass almost irrespective of the cycle length, although for $\;\tcyc\le 100\;\Myr$ we find that the individual shocks occur so regularly that they prevent the gas from cooling. For long cycle lengths $\tcyc \ge 2000\,\Myr$ we also find a slight decline in our cool mass fraction which is caused by the cool outflows being slower in this regime (so that a smaller proportion pass the velocity cut).

We can also generate large cool-gas fractions in a few cases that we classified as chimneys (middle-top part of Fig.~\ref{dutyvscycledots}) at $f_{\mathrm{cyc}}=0.5$, with $t_{\mathrm{cyc}} = 3000$ or  $2500$. In these cases, galaxies spend half their time injecting hot gas into the halo. This means that the halo is left undisturbed for a period of time $>\mathrm{1\; Gyr}$ which is long enough for even relatively diffuse gas at $\sim 10^4 \,\Msol\,\kpc^{-3}$ to cool. As seen in Fig.~\ref{illustration}, there is gas at this density in the regions around our chimneys. However for the cases where the cycle fraction reaches $f_{\mathrm{cyc}}=0.75$ (towards the top right of the figure),  the period for which the halo is undisturbed by incoming material becomes less than the cooling time of approximately 1 Gyr, and the cool gas fraction declines.

By contrast in the low-$f_{\mathrm{cyc}}$ regime, the regular shock fronts between new gas and the existing halo always reach high densities ($\gg \,10^{4}\;\mathrm{M_{\odot}\,kpc}^{-3}$), bringing the cooling time down to a few hundred Myr (depending on the exact temperature and density) and so allowing efficient cold gas formation.

{ Figure 6 highlights the existence of a transition between $f_{\mathrm{cyc}}=0.15$ and 0.25 which falls in between these cases and is unable to efficiently produce cool gas.  In these cases, gas behind the shock is hotter and denser than its environment. In fact we find that, during the ``off'', relatively undisturbed phase, a portion of the halo does cool; however, it is slow-moving and starts to infall; it is therefore not counted since we made a cut to include only gas particles travelling faster than 50 km $\mathrm{s^{-1}}$. The gas that is outflowing in these cases remains hot ($10^5$ K) and with low density ($\sim$ 80 $\mathrm{M_{\odot}\,kpc^{-3}}$). 
}

	
\section{Conclusions}
\label{conclusions}

In this paper we have considered the possibility that the cool gas material observed in outflows in the circumgalactic medium is a consequence of in situ cooling \citep{2016MNRAS.455.1830T}, using the {\sc Ramses} code. We do not simulate the disk in our galaxies; instead we manually inject gas moving into the base of a box representing the halo \citep[see also][]{1999ApJ...513..142M}. We started by finding and testing equilibrium inflows to be sure that the effects we observe are a result of outflows rather than the choice of initial conditions; see Eqs. \eqref{eqn:u} -- \eqref{eqn:Temp}. We used a fixed potential corresponding to a $50\;\km\,\s^{-1}$ virial velocity dwarf galaxy. The cooling function $\dot Q_{\mathrm{cool}}$ we adopt is suitable for primordial gas as implemented by {\sc Ramses}, which underestimates the true cooling rates and so leads us to conservative conclusions.

We modified {\sc Ramses} to inject a time-varying flow into the bottom of the computational domain, according to a set of parameters summarised in Table \ref{on_off}. Our particular focus was on the role that varying star formation rates in the galaxy could have on the evolution of outflows as they traverse the halo.  From the complete set of parameters characterising outflows we therefore varied two: the overall star formation cycle length, $\tcyc$, and the fraction of that time spent pumping gas into the CGM, $\fcyc$. We varied these while keeping the total energy injection and mass loading constant. 
We found a close connection between the parameters and the overall nature of the outflows' traversal of the CGM (Fig.~\ref{pattern}). This in turn has a strong effect on the multiphase nature of the outflows.

The amount of cooler $T<5 \times 10^4\,\K$ gas present in the outflow varies strongly over the course of a cycle (Fig.~\ref{coldgas}). Cool material is typically able to form as a shock propagates outwards provided that no hot material is being injected behind the shock. This leads to the time-averaged cool mass fraction depending on both $\tcyc$ and $\fcyc$ (Fig.~\ref{dutyvscycledots}). There are two regimes in which we obtain large cool mass fractions: the first has a small $\fcyc$, corresponding to a rapidly fluctuating star formation rate. Provided $\tcyc$ is greater than a few hundred $\Myr$, the successive shocks do not join up into a coherent flow and the strong time variability triggers waves of effective cooling that travel through the CGM. The second approach is to leave a long period $\gsim 1\;\Gyr$ between successive star formation epochs; in this case cool gas is able to form in the turbulent halo left behind when ejection from the disk shuts off. Our results suggest that steady flow or single-burst solutions with cooling \citep{2016MNRAS.455.1830T} form a lower bound on the amount of in situ cooling to be expected in realistic galaxies with time-varying feedback.

Our interest in cool gas is motivated by observational results that show the presence of a cool phase even at large distances from galactic centres \citep[e.g.,][]{Nielsen2013}, which is hard to explain in entrainment scenarios \citep{2015arXiv150701951Z}. However it would be premature to compare our highly idealised study directly to observations. { To enable our large parameter study, we had to restrict ourselves to two-dimensional solutions; the detailed behaviour in three dimensions could differ significantly. Other neglected aspects of the problem include the enhanced cooling rates from metal enrichment and the effects of thermal conduction and magnetic fields. Furthermore a realistic cosmological environment is far more complex than the uniform inflow solution that forms our initial conditions.  In terms of the cooling rates, neglecting metals leads to a conservative bound; i.e., more realistic simulations may be able to form cold clouds more easily than our work suggests.}

We hope to use our results to interpret the CGM around ab initio cosmological galaxy formation simulations. Some feedback algorithms enforce relatively steady-state star formation \cite[e.g.,][]{2008MNRAS.387.1431D,2008MNRAS.387..577O,2014MNRAS.444.1518V,2015MNRAS.446..521S} whereas others lead to strong bursts \cite[e.g.,][]{2014MNRAS.445..581H,2015MNRAS.453.3499K} and the importance of this distinction for galactic dynamics has already been established \citep{pontzen}. In future work we will study what role in situ cooling plays in these different scenarios, and make the link to observational constraints on the rich phenomenology of the CGM.

\section*{Acknowledgments}
We thank the referee, Evan Scannapieco, for a helpful report with many useful suggestions. TS acknowledges support via studentships from CONACyT (Mexico) and the University College London.  AP is supported by the Royal Society. HVP was supported by the European Research Council under the European Community's Seventh Framework Programme (FP7/2007- 2013) / ERC grant agreement no 306478-CosmicDawn. The simulations were analysed using {\sc Pynbody} \citep{2013ascl.soft05002P}. This work used the DiRAC Complexity system, operated by the University of Leicester IT Services, which forms part of the STFC DiRAC HPC Facility ({\tt www.dirac.ac.uk}). This equipment is funded by BIS National E-Infrastructure capital grant ST/K000373/1 and STFC DiRAC Operations grant ST/K0003259/1. DiRAC is part of the National E-Infrastructure.

\bibliographystyle{mn2e}
\bibliography{outflows.bib}

\end{document}